\documentclass[
reprint,
groupedaddress,
amsmath,amssymb,
prmaterials,
floatfix,
superscriptaddress,
]{revtex4-2}

\usepackage{amsmath}
\usepackage{bm}
\usepackage{amsfonts}
\usepackage{graphicx}
\usepackage{amssymb}
\usepackage[caption=false]{subfig}
\usepackage{hyperref}
\usepackage{cleveref}
\usepackage{siunitx}
\usepackage[T1]{fontenc}
\usepackage{libertine}

\DeclareSIUnit\angstrom{\text{\AA}}

\begin{document}

\title{Chemo-mechanics in alloy phase stability}
\author{Sesha Sai Behara}
\affiliation{Materials Department, University of California, Santa Barbara}
\date{\today}
\author{John C. Thomas}
\affiliation{Materials Department, University of California, Santa Barbara}
\date{\today}
\author{Brian Puchala}
\affiliation{Department of Materials Science and Engineering, University of Michigan, Ann Arbor}
\date{\today}
\author{Anton Van der Ven}
\email{avdv@ucsb.edu}
\affiliation{Materials Department, University of California, Santa Barbara}
\date{October 30, 2023}

\begin{abstract}

We describe a first-principles statistical mechanics method to calculate the free energies of crystalline alloys that depend on temperature, composition, and strain. 
The approach relies on an extension of the alloy cluster expansion to include an explicit dependence on homogeneous strain in addition to site occupation variables that track the degree of chemical ordering. 
The method is applied to the Si-Ge binary alloy and is used to calculate free energies that describe phase stability under arbitrary epitaxial constraints. 
We find that while the incoherent phase diagram (in which coexisting phases are not affected by coherency constraints) hosts a miscibility gap, coherent phase equilibrium predicts ordering and negative enthalpies of mixing. 
Instead of chemical instability, the chemo-mechanical free energy exhibits instabilities along directions that couple the composition of the alloy with a volumetric strain order parameter. 
This has fundamental implications for phase field models of spinodal decomposition as it indicates the importance of gradient energy coefficients that couple gradients in composition with gradients in strain. 

\end{abstract}

\maketitle
\section{Introduction}\label{sec:intro}

Strain is increasingly used as an explicit thermodynamic boundary condition to manipulate the properties of materials. 
Most materials, however, exhibit some chemical complexity as one or more of their sublattices can host multiple chemical species that may order at low temperatures and form disordered solid solutions at high temperatures. 
There is, therefore, the possibility of important and interesting chemo-mechanical couplings that can emerge, whereby changes in composition can affect elastic properties and changes in strain can affect chemical interactions. 

Early work by Larche and Cahn \cite{larche1978nonlinear,larche1978thermochemical,larche1985overview} and Voorhees and Johnson \cite{voorhees2004thermodynamics} set up a phenomenological framework that rigorously integrates chemical alloy thermodynamics with elasticity and showed that coherency constraints can have a pronounced effect on phase stability and the topology of phase diagrams. 
The effect of epitaxial constraints on phase stability has been found to be especially strong in semiconductor alloys. 
Many semiconductor alloys that exhibit miscibility gaps in their equilibrium temperature versus composition phase diagrams can form ordered compounds when epitaxially constrained.\cite{zunger1989structural,martins1986stability,wood1989epitaxial}
More generally, the coupling between strain, composition and degree of ordering in epitaxially grown semiconductor alloys affects phase stability, solubility and their resulting optoelectronic properties.\cite{Donner2011, Kuech2013, kuech2016growth, Holder2017, Jacobsen2012, stevens2019, Guan2020, Schnepf2020, Cordell2022} 
Coherency constraints also affect the electrochemical properties of battery electrode materials as they undergo phase transitions \cite{van2009role} as well as the shapes and compositions of coherent precipitates in metal alloys
\cite{huh1993effect,doi1996elasticity,natarajan2016early,natarajan2017unified,dewitt2017misfit,solomon2019stability}
and the electrocatalysis activity and ferroelectric properties of perovskite thin films.\cite{Hwang2019}

First-principles statistical mechanics treatments of chemical order-disorder phenomena usually neglect strain as a thermodynamically controlled boundary condition. 
The cluster expansion formalism introduced by Sanchez et al.\cite{sanchez1984generalized} is a mathematically rigorous surrogate model that enables an efficient and compact description of the energy of a crystal as a function of the degree of ordering among its different chemical constituents (or of the orientational arrangements of magnetic moments \cite{drautz2004spin,decolvenaere2019modeling,kitchaev2020mapping} or molecules \cite{thomas2018hamiltonians}).
An alloy cluster expansion that has been trained to first-principles electronic structure calculations can be used to predict alloy thermodynamic properties at finite temperature with mean field or Monte Carlo methods.\cite{de1994cluster,van2018first} 
The cluster expansion surrogate model, though, is either used to parameterize the configurational energy of an alloy at constant volume \cite{sanchez2017foundations} or in the fully relaxed state \cite{wolverton1995ising,zunger1994first} (i.e. at zero pressure) but is not able to generate information about the dependence of thermodynamic properties on strain. 
Past treatments of the effect of strain on chemical equilibrium in crystalline alloys using the mixed basis cluster expansion of Laks et al.\cite{laks1992efficient}, account for the long-range interactions that arise from coherency strains, but treat strain implicitly and not as an explicit degree of freedom. 
The cluster expansion approach has also been extended to describe the dependence of tensor properties such as elastic moduli on the chemical degree of ordering within a crystal.\cite{vandewalleNatMat,liu2005structure}
While this makes it possible to calculate elastic moduli as a function of temperature and composition, it only provides derivative properties of a full chemo-mechanical free energy description of a solid.

In this contribution, we describe a strain plus configuration cluster expansion that has an explicit dependence on both the chemical order and the homogeneous strain of an alloy.
The approach enables the calculation of free energies that are functions of temperature, composition, and strain, thereby rigorously encapsulating the chemo-mechanical couplings that affect all thermodynamic properties that can be derived from free energy, including chemical potentials, elastic moduli, and phase stability. 
The free energies are also well suited to feed into phase field models of microstructure evolution, coherent spinodal decomposition, and precipitation due to coherent ordering reactions.\cite{chen2002phase,rudraraju2016mechanochemical,natarajan2017symmetry,van2018first,teichert2019machine,teichert2020scale} 
To illustrate the approach, we calculate chemo-mechanical thermodynamic properties of the Si-Ge binary alloy, which can form a disordered solid solution over the sites of the diamond parent crystal structure.

\section{Methods}

\subsection{Chemical and strain descriptors}

Chemical degrees of freedom in a crystal can be tracked with occupation variables assigned to each crystal site. 
For a binary A-B alloy, the occupation variables, $\sigma_l$, attached to site $l$ can, for example, take a value of +1 if occupied by A and -1 if occupied by B. 
The chemical arrangement of the whole crystal is then uniquely specified with a vector of occupation variables, $\vec{\sigma}=(\sigma_1,\dots,\sigma_l,\dots,\sigma_N)$, where $N$ is the total number of sites that can be decorated with A and B atoms. 

The degree with which a crystal is homogeneously deformed can be tracked with six independent metrics of strain, $E_{xx}$, $E_{yy}$, $E_{zz}$, $E_{yz}$, $E_{zx}$, and $E_{xy}$. 
These measure the deformation of a crystal relative to its dimensions in a reference state and are defined to be independent of rigid rotations \cite{thomas2017exploration}. 
The numerical values of each strain component depend on the chosen orientation of the Cartesian coordinate system. 
The optimal orientation is usually guided by symmetry considerations of the reference state used to define strain. 
For a crystal that has cubic symmetry in the unstrained state, for example, the Cartesian axes are usually chosen to be parallel to the lattice vectors of the cubic unit cell.

There are several definitions of finite strain. 
The general approach of describing finite strains starts with the deformation gradient, a $3\times 3$ matrix $\mathbf{F}$, that for homogeneous deformations of a crystal relates the lattice vectors of the reference crystal, $\vec{a}$, $\vec{b}$ and $\vec{c}$, to the lattice vectors of the deformed crystal, $\vec{A}$, $\vec{B}$ and $\vec{C}$, according to
\begin{equation}
    [\vec{A}, \vec{B}, \vec{C}]=\mathbf{F}[\vec{a}, \vec{b}, \vec{c}]
\end{equation}
In this relationship, the Cartesian coordinates of the lattice vectors are collected as columns in $3\times3$ matrices represented by $[\vec{A}, \vec{B}, \vec{C}]$ and $[\vec{a}, \vec{b}, \vec{c}]$. 
While the deformation gradient, $\mathbf{F}$, contains information about the degree to which the crystal has been deformed, it is also affected by any rigid rotation of the deformed crystal. 
To eliminate the dependence on rigid rotations, metrics of strain are defined in terms of the rotation invariant quantity $\mathbf{F}^{\mathsf{T}}\mathbf{F}$. 
The Green-Lagrange strain, for example, takes the form $\mathbf{E}=1/2(\mathbf{F}^{\mathsf{T}}\mathbf{F}-\mathbf{I})$, where $\mathbf{I}$ is the identity matrix, while the Hencky strain is defined as $\mathbf{E}=1/2\ln(\mathbf{F}^{\mathsf{T}}\mathbf{F})$.

It is often convenient to work with symmetry-adapted strain order parameters instead of the Cartesian strains directly since order parameters provide more insight about the symmetries that are broken due to a particular strain. 
When the reference state of the crystal has cubic symmetry, an especially descriptive set of strain order parameters can be defined according to
\begin{equation}
    \begin{pmatrix}
    e_1 \\
    e_2 \\
    e_3 \\
    e_4 \\
    e_5 \\
    e_6
    \end{pmatrix}
    =
     \begin{pmatrix}
     \frac{(E_{xx}+E_{yy}+E_{zz})}{\sqrt{3}} \\
     \frac{(E_{xx}-E_{yy})}{\sqrt{2}} \\
     \frac{(2E_{zz}-E_{xx}-E_{yy})}{\sqrt{6}} \\
     \sqrt{2}E_{yz} \\
     \sqrt{2}E_{xz} \\
     \sqrt{2}E_{xy}
     
     \end{pmatrix}
     \label{eq:cubic_sops}
\end{equation}

The first strain order parameter, $e_1$, is invariant to symmetry and measures volumetric changes of the crystal as a result of deformation. 
The next two order parameters, $e_2$ and $e_3$, together form an irreducible subspace when the reference crystal has cubic symmetry and describe the tetragonal and orthorhombic distortions of the cubic reference.
The last three strain order parameters, $e_4$, $e_5$ and $e_6$, are shear strains.
One of the advantages of working with the Hencky strain metric is that the first strain order parameter, $e_1$, is exclusively determined by the relative change in volume of the crystal and any variation of the other strain order parameters, $e_2,\dots,e_6$, at constant $e_1$ corresponds to symmetry breaking deformations at constant volume.\cite{thomas2017exploration}

\begin{table*}[]
    \centering
    \caption{The first 23 prototype cluster basis functions of a configuration plus strain cluster expansion for the diamond parent crystal structure. The $e_1$, $e_2$, $e_3$, $e_4$, $e_5$ and $e_6$ variables are the strain order parameters defined by Eq. \ref{eq:cubic_sops}. The $\sigma$ represents the occupation variables that are assigned to each crystal site. These take the values of -1 or +1 depending on the occupant of the site (i.e. Si or Ge) for the Ge$_{1-x}$Si$_x$ alloy. The first eleven basis functions have only a strain dependence. The next five basis functions correspond to point cluster terms. Basis functions $\Phi_{16}$ to $\Phi_{18}$ are prototype basis cluster functions coupling occupants on the nearest neighbor pair cluster with indices $i$ and $j$, while basis functions $\Phi_{19}$ to $\Phi_{22}$ couple occupants on the next nearest neighbor pair cluster of the diamond parent crystal. The specific form of the strain dependent polynomials of each cluster basis function depends on the orientation of the cluster relative to the Cartesian coordinate system.}
    \begin{tabular}{c c}
        \hline
        \hline
        Cluster function notation & Prototype basis function \\
        \hline
         $\Phi_{0}$ & 1\\
            $\Phi_{1}$ & $e_1$\\
            $\Phi_{2}$ & $e_1^{2}$\\
            $\Phi_{3}$ & $\sqrt{1/2}(e_2^{2} +e_3^{2} )$ \\
            $\Phi_{4}$ & $\sqrt{1/3}(e_4^{2} +e_5^{2} +e_6^{2} )$\\
            $\Phi_{5}$ & $e_1^{3}$\\
            $\Phi_{6}$ & $\sqrt{3/2}(e_1e_2^{2} +e_1e_3^{2} )$\\
            $\Phi_{7}$ & $3/2(e_2^{2} e_3-1/3e_3^{3} )$\\
            $\Phi_{8}$ & $(e_1e_4^{2} +e_1e_5^{2} +e_1e_6^{2} )$\\
            $\Phi_{9}$ & $\sqrt{3/4}(e_2e_4^{2} -e_2e_5^{2} -\sqrt{1/3}e_3e_4^{2} -\sqrt{1/3}e_3e_5^{2} +\sqrt{4/3}e_3e_6^{2} )$\\
            $\Phi_{10}$ & $\sqrt{6}e_4e_5e_6$\\
        \hline
        $\Phi_{11}$ & $\sigma_i$\\
            $\Phi_{12}$ & $\sigma_i e_1$\\
            $\Phi_{13}$ & $\sigma_i e_1^{2} $\\
            $\Phi_{14}$ & $\sigma_i\sqrt{1/2} (e_2^{2} +e_3^{2} )$\\
            $\Phi_{15}$ & $\sigma_i \sqrt{1/3}(e_4^{2}+e_5^{2} +e_6^{2})$\\
        \hline
        $\Phi_{16}$ & $\sigma_i\sigma_j$\\
        $\Phi_{17}$ & $\sigma_i\sigma_je_1$\\
        $\Phi_{18}$ & $\sigma_i\sigma_j\sqrt{1/3}(e_4-e_5-e_6)$\\
        \hline
         $\Phi_{19}$ & $\sigma_i\sigma_j$\\
         $\Phi_{20}$ & $\sigma_i\sigma_je_1$\\
        $\Phi_{21}$ & $\sigma_i\sigma_j\sqrt{3/4}(e_2-\sqrt{1/3}e_3)$\\
            $\Phi_{22}$ & $\sigma_i\sigma_je_4$\\
            \hline
            \hline
    \end{tabular}
    \label{tab:basis_functions}
\end{table*}

\subsection{Coupling chemical and strain degrees of freedom}

In formulating a strain plus configuration cluster expansion, we consider a binary A-B alloy, but the approach can be generalized to an arbitrary number of chemical constituents.\cite{thomas2023}
Following the derivation of Sanchez et al \cite{sanchez1984generalized}, the cluster expansion expression for the energy of a crystal can be generalized to explicitly depend on strain, $\vec{e}$, in addition to configuration, $\vec{\sigma}$, according to
\begin{equation}
    E(\vec{e},\vec{\sigma})= \sum_{\alpha}\sum_{n_{\alpha}}V_{\alpha}^{n_{\alpha}}\Phi_{\alpha}^{n_{\alpha}}(\vec{e},\vec{\sigma})
    \label{eq:strain_cluster_expansion1}
\end{equation}
where 
\begin{equation}
    \Phi_{\alpha}^{n_{\alpha}}(\vec{e},\vec{\sigma})=\lambda_{\alpha}^{n_{\alpha}}(\vec{e})\phi_{\alpha}(\vec{\sigma})
\end{equation}
are cluster basis functions. 
The index $\alpha$ labels clusters of sites in the crystal, such as point, pair, triplet, etc. clusters.
The sum extends over all possible $2^{N}$ clusters of sites within a crystal, where $N$ is the total number of sites in the crystal that can be occupied by A or B atoms. 
The cluster basis functions $\Phi_{\alpha}^{n_{\alpha}}(\vec{e},\vec{\sigma})$ are products of functions of strain, $\lambda_{\alpha}^{n_{\alpha}}(\vec{e})$ with polynomials of the occupation variables defined as
\begin{equation}
    \phi_{\alpha}(\vec{\sigma})=\prod_{l\in \alpha}\sigma_l.
\end{equation}
The $\phi_{\alpha}(\vec{\sigma})$ are the cluster basis functions of the original binary alloy cluster expansion introduced by Sanchez et al. \cite{sanchez1984generalized} and are simply a product of occupation variables belonging to the sites of the cluster $\alpha$.
The strain basis functions, $\lambda_{\alpha}^{n_{\alpha}}(\vec{e})$, can similarly be represented as polynomials of the six strain variables, $\vec{e}$. 
The $n_{\alpha}$ is an integer index that tracks distinct polynomials of strain.
Finally, $V_{\alpha}^{n_{\alpha}}$ are expansion coefficients that embed the specific chemical and elastic properties of the alloyed crystal and are referred to as effective cluster interactions (ECI). 

The symmetry of the undecorated and unstrained reference parent crystal imposes constraints on the expansion coefficients, $V_{\alpha}^{n_{\alpha}}$, and the functional form of the basis functions, $\Phi_{\alpha}^{n_{\alpha}}(\vec{e},\vec{\sigma})$. 
This is because any pair of chemical orderings and/or strain states, $(\vec{e},\vec{\sigma})$ and $(\vec{e}',\vec{\sigma}')$, that can be mapped onto each other upon the application of a space group symmetry operation, $\hat{S}$, of the parent crystal structure must have the same energy. 

In generating a suitable basis that satisfies all symmetry constraints, it is useful to rely on several group theoretical concepts related to crystal symmetry and clusters. 
The first is an orbit of clusters, defined as the collection of all clusters of a particular type that are equivalent by symmetry. 
By starting with a prototype cluster, $\alpha$, such as the nearest neighbor pair, all other clusters belonging to its orbit, $\Omega_{\alpha}$, can be generated upon the application of the space group operations of the parent crystal to the prototype cluster $\alpha$. 
Each distinct cluster type (e.g. point cluster, nearest neighbor pair cluster, second nearest neighbor pair cluster, etc.) has a corresponding orbit that collects all symmetrically equivalent clusters of that type.
Another useful concept is a cluster group. 
A prototype cluster, $\alpha$, of the cluster orbit $\Omega_{\alpha}$ has a cluster group, $G_{\alpha}$, consisting of a subset of space group operations of the parent crystal that maps the cluster $\alpha$ onto itself.

A set of cluster basis functions that ensure the invariance of the energy of the crystal to the symmetry of the parent crystal structure can be generated as follows.  
For a prototype cluster, $\alpha$, of each cluster orbit, $\Omega_{\alpha}$, generate a starting basis of strain polynomials in terms of monomials of each strain variable, $e_i$, of the form
\begin{equation}
    \eta^{\vec{p}}_{\alpha}(\vec{e})=\prod_{i=1}^{6}e_{i}^{p_i}
    \label{eq:starting_strain_basis}
\end{equation}
where $\vec{p}=(p_1,p_2,\dots,p_6)$ and where each individual $p_i$ is a positive integer ranging from 0 to a maximum polynomial order. 
In this expression, each $e_i^{p_i}$ is a monomial of the strain variable $e_i$ to the $p_i^{th}$ power.
The strain polynomials, $\eta^{\vec{p}}_{\alpha}(\vec{e})$ can be made invariant to the cluster group, $G_{\alpha}$, upon the application of the Reynolds operator, defined as 
\begin{equation}
    \lambda_{\alpha}^{n_{\alpha}}(\vec{e})=\frac{1}{|G_{\alpha}|}\sum_{\hat{S}\in G_{\alpha}} \hat{S}[\eta^{\vec{p}}_{\alpha}(\vec{e})]
\end{equation}
where $|G_{\alpha}|$ corresponds to the number of symmetry operations in the cluster group $G_{\alpha}$ and $\hat{S}[\eta^{\vec{p}}_{\alpha}(\vec{e})]$ represents the application of the symmetry operation $\hat{S}$ to the function $\eta^{\vec{p}}_{\alpha}$. 
The application of the Reynolds operator to the starting basis polynomials generated by enumerating all possible non-negative integers $p_i$ in Eq. \ref{eq:starting_strain_basis} produces a set of symmetry invariant polynomials.\cite{thomas2017exploration} 
Some of these polynomials may be duplicates or linear combinations of other symmetry invariant polynomials. 
After discarding the linearly dependent polynomials, a set of symmetry invariant polynomials, $\lambda_{\alpha}^{n_{\alpha}}(\vec{e})$, can be collected, each labeled with a distinct integer index $n_{\alpha}$. 
More details about the application of symmetry to strain variables and to polynomials of strain, the Reynolds operator, and the determination of a linearly independent set of symmetry invariant polynomials can be found in Thomas and Van der Ven \cite{thomas2017exploration}. 

Once a set of symmetry invariant cluster basis functions, $\Phi_{\alpha}^{n_{\alpha}}(\vec{e},\vec{\sigma})$ with $n_{\alpha}= 1,\dots, n_{\alpha}^{max}$ for a prototype cluster $\alpha$, have been generated, all other symmetrically equivalent cluster basis functions within the crystal, $\Phi_{\beta}^{n_{\alpha}}(\vec{e},\vec{\sigma})$ with $\beta \in \Omega_{\alpha}$, can be generated upon the application of space group operations to the prototype cluster basis functions $\Phi_{\alpha}^{n_{\alpha}}(\vec{e},\vec{\sigma})$. 
The resulting symmetrically equivalent basis functions can be collected in an orbit of basis functions labeled $\Omega_{\alpha}^{n_{\alpha}}$. 
The basis functions belonging to the same basis function orbit, $\Omega_{\alpha}^{n_{\alpha}}$, all share the same expansion coefficient $V_{\alpha}^{n_{\alpha}}$ in a symmetry invariant cluster expansion, making it possible to rewrite the strain plus configuration cluster expansion of Eq. \ref{eq:strain_cluster_expansion1} as  
\begin{equation}
    E(\vec{e},\vec{\sigma})= \sum_{\Omega_{\alpha}^{n_{\alpha}}}V_{\alpha}^{n_{\alpha}}\left(\sum_{\beta \in \Omega_{\alpha}}\Phi_{\beta}^{n_{\alpha}}(\vec{e},\vec{\sigma})\right)
    \label{eq:strain_cluster_expansion2}
\end{equation}
In this form, the outer sum is over distinct orbits of cluster basis functions, $\Omega_{\alpha}^{n_{\alpha}}$, and the inner sum is over all clusters that are symmetrically equivalent to the prototype cluster $\alpha$.

\Cref{tab:basis_functions} lists the first 23 terms of a strain plus configurational cluster expansion for a diamond parent crystal structure.
The first term in the expansion, Eq. \ref{eq:strain_cluster_expansion2}, corresponds to the empty cluster, $\alpha$ = $\varnothing$, and has exclusively a strain dependence.
There are also other basis functions in the expansion that only depend on occupation variables and are strain independent. 
Most of the basis functions, though, depend on both occupation variables and a subset of strain variables. 

The strain plus configuration cluster expansion of Eq. \ref{eq:strain_cluster_expansion2} is for the energy of the whole crystal and is an extensive quantity. 
This expression can be normalized by the number of unit cells of the crystal, $N_{u}$, upon the introduction of correlation functions defined as 
\begin{equation}
    \xi_{\alpha}^{n_{\alpha}}(\vec{e},\vec{\sigma})= \frac{1}{N_{u}m_{\alpha,n_{\alpha}}}\sum_{\beta \in \Omega_{\alpha}}\Phi_{\beta}^{n_{\alpha}}(\vec{e},\vec{\sigma})
\end{equation}
where $m_{\alpha,n_{\alpha}}$ is the multiplicity of the cluster basis function, $\Phi_{\alpha}^{n_{\alpha}}(\vec{e},\vec{\sigma})$
, per unit cell (i.e. $m_{\alpha,n_{\alpha}} = |\Omega_{\alpha}^{n_{\alpha}}|/N_{u}$). 
The energy per unit cell, $\epsilon(\vec{e},\vec{\sigma})=E(\vec{e},\vec{\sigma})/N_{u}$, then becomes
\begin{equation}
    \epsilon(\vec{e},\vec{\sigma})= \sum_{\Omega_{\alpha}^{n_{\alpha}}}V_{\alpha}^{n_{\alpha}}m_{\alpha,n_{\alpha}}\xi_{\alpha}^{n_{\alpha}}(\vec{e},\vec{\sigma})
\end{equation}
Expressed in this manner, an approach to parameterize the expansion coefficients to the results of first-principles electronic structure training data becomes evident. 
In practice, the strain plus configuration cluster expansion must be truncated. 
The energies of a large number of strain and chemical configurations can be calculated with a first-principles electronic structure method and used to determine the expansion coefficients $V_{\alpha}^{n_{\alpha}}$ of a truncated cluster expansion using one of many possible machine-learning regression techniques, such as simple least squares, LASSO and ridge regression.\cite{Bishop2006,hart2005evolutionary,mueller2009bayesian,nelson2013cluster,nelson2013compressive,ober2023thermodynamically}

The CASM software package \cite{puchala2023casm,puchala2023casmmonte,thomas2023} is capable of algorithmically generating a strain plus configurational cluster expansion for any parent crystal structure. 
It also has capabilities to enumerate structural models with symmetrically distinct strains and configurations, $(\vec{e}, \vec{\sigma})$, and to calculate the correlation functions for each cluster basis function in the configuration $(\vec{e}, \vec{\sigma})$.

\subsection{First-principles electronic structure calculations to parameterize the cluster expansion}

The CASM software package was used to enumerate crystallographic models having different arrangements of Si and Ge over the sites of the diamond parent crystal structure and to enumerate different states of strain of each ordered configuration.\cite{puchala2023casm}
The formation energy of each enumerated structure was calculated using the Vienna Ab-initio Simulation Package (VASP) \cite{kresse1993ab, kresse1996efficiency, kresse1996efficient}. 
The density functional theory (DFT) calculations were performed within the Generalized Gradient Approximation (GGA) using the PBE exchange-correlation functional \cite{perdew1996generalized}. 
The projector augmented wave (PAW) method \cite{blochl1994projector} as implemented within VASP with Ge$_d$ and Si pseudopotentials was used.  
A plane-wave energy cutoff of 400eV was used for all the calculations. 
An automatic $\Gamma$-centered \textit{k}-point mesh was generated for all the structures using a length parameter $R_k$ of 35. 
A value of $10^{-5}$ eV was used as an electronic convergence criterion. 
To achieve geometric convergence, a cutoff value of 0.02 eV/\si{\angstrom} was used for forces on all atoms.

\begin{figure}
    \subfloat[]{\includegraphics[width=8cm]{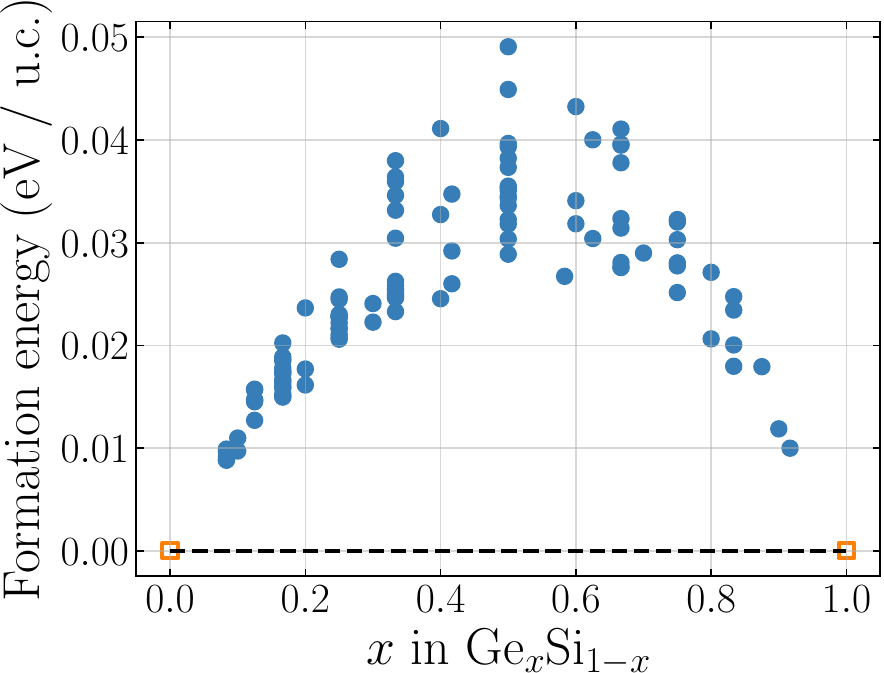}\label{fig:SiGe_relaxedhull}}\\
    \subfloat[]{\includegraphics[width=8cm]{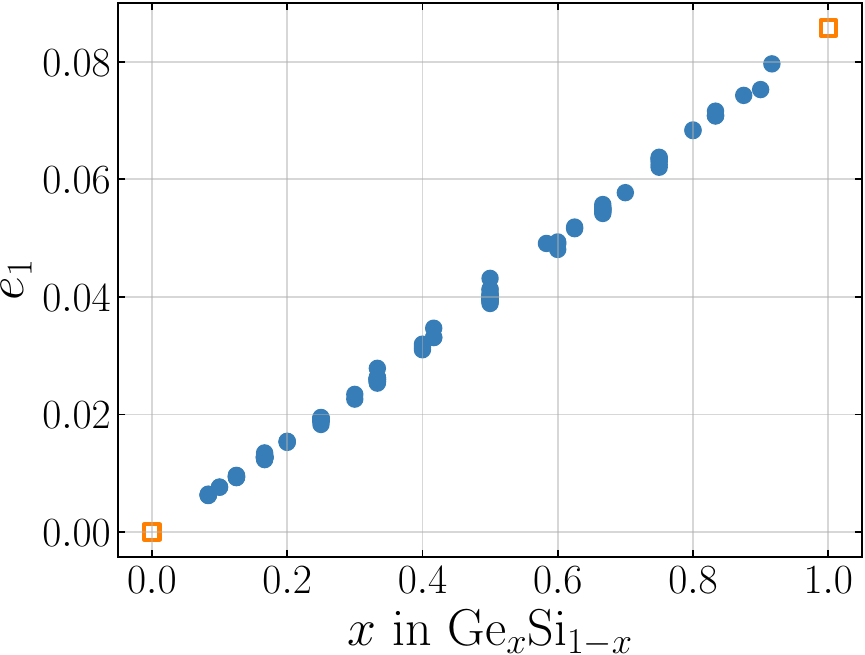}\label{fig:e1vsx}} 
    \caption{(a) Formation energies of fully relaxed Si-Ge configurations. Si and Ge in the fully relaxed diamond parent crystal structure were used as reference states. (b) $e_1$ strain order parameter for fully relaxed Si-Ge configurations with pure Si as the reference state}
    \label{fig:SiGe_dft}
\end{figure}

\section{Results}

The strain plus configuration cluster expansion approach was applied to study finite temperature phase stability and chemical disorder in the Si-Ge alloy. 
Both Si and Ge adopt the diamond crystal structure and form high temperature solid solutions. 
The lattice parameters of pure Si differ substantially from those of Ge, indicating that coherency strains play an important role in determining phase stability.

\begin{figure}[htbp]
    \subfloat[]{\includegraphics[width=6.5cm]{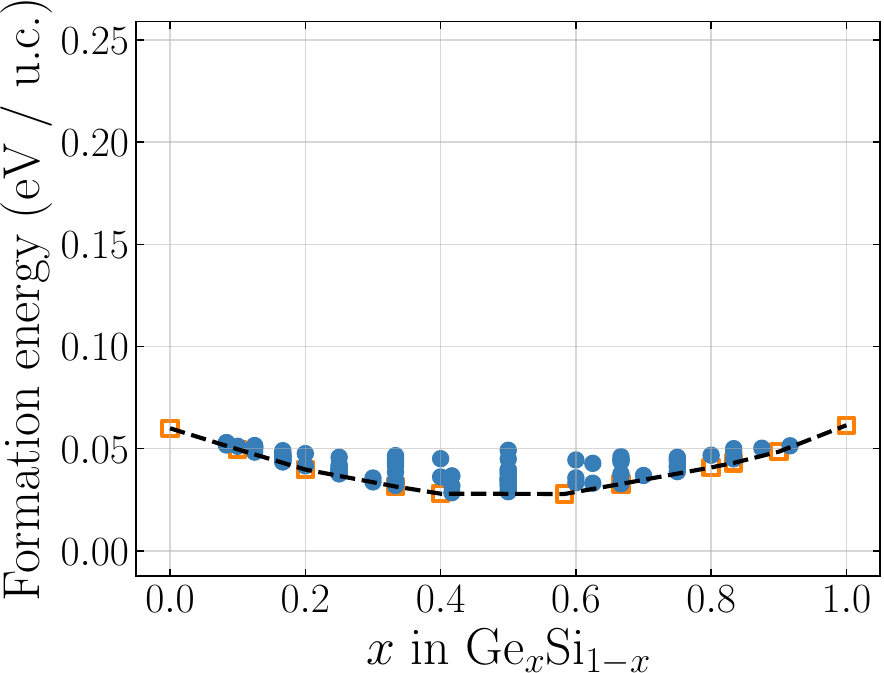}\label{fig:SiGeHullatx0.5}}\\
    \subfloat[]{\includegraphics[width=6.5cm]{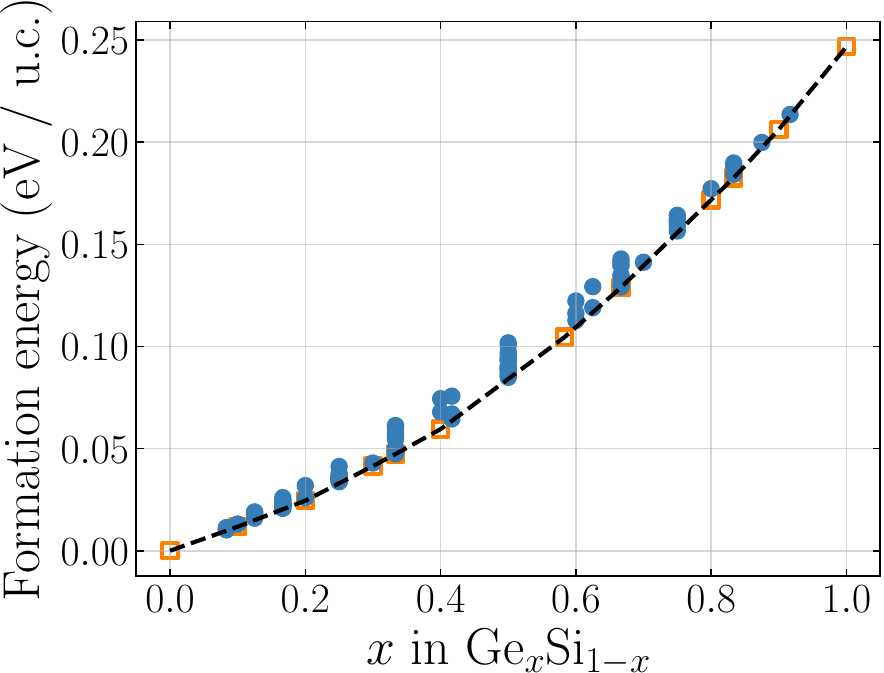}\label{fig:SiGeHullatpureSi}}\\
    \subfloat[]{\includegraphics[width=6.5cm]{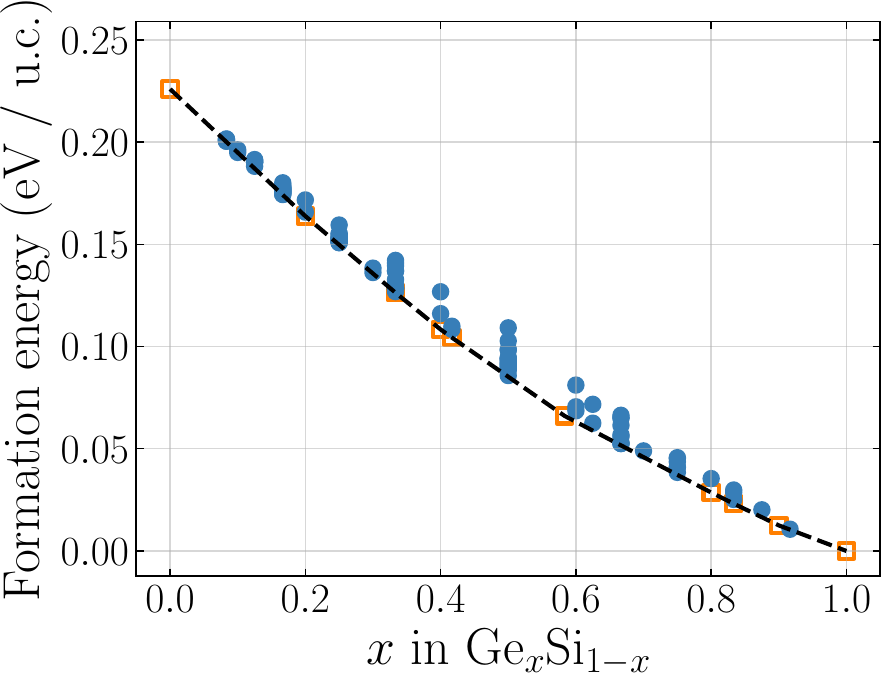}\label{fig:SiGeHullatpureGe}}
    {
    \caption{Formation energies of various Si-Ge configurations at fixed equilibrium volumes of (a) a Si-Ge alloy at $x = 0.5$ (b) pure Si (c) pure Ge. The dotted black line represents the thermodynamic convex hull whereas the orange squares represent the ground state configurations that fall on the convex hull.} 
    \label{fig:hullsatdifferente1s}
    }
\end{figure}

\subsection{Formation energies, relaxation strains, and cluster expansion}
\Cref{fig:SiGe_relaxedhull} shows the formation energies of 119 fully relaxed Si-Ge configurations on the sites of the diamond parent crystal as a function of the Ge composition. 
The reference states are pure Si and pure Ge in the diamond crystal structure. 
The positive formation energies signify a thermodynamic tendency for phase separation and the existence of a miscibility gap in the equilibrium temperature versus composition phase diagram at constant pressure. 
\Cref{fig:e1vsx} shows the values of the $e_1$ strain order parameter for each fully relaxed Si-Ge configuration, calculated using the lattice parameter of pure Si in the diamond crystal as the reference state. 
The plot shows a very strong and almost linear volume change with increasing Ge concentration, but a very weak dependence of the volume on the nature of the Si-Ge ordering as there is very little variation in $e_1$ values at each composition.

The strong dependence of the equilibrium volume of the Si-Ge alloy with Ge concentration indicates that coherency constraints should have a significant effect on phase stability. 
This is manifested in \Cref{fig:SiGeHullatx0.5}, which shows the formation energies of the 119 Si-Ge configurations calculated at a volume that is fixed to the approximate equilibrium volume of the alloy at $x=0.5$. 
While the unit cell dimensions of each configuration were held fixed, all internal atomic coordinates were allowed to relax in these calculations. 
\Cref{fig:SiGeHullatx0.5} shows that the constraint of a constant volume set at its x=0.5 value severely penalizes the energies of the Si rich and the Ge rich configurations. 
Instead of exhibiting a driving force for phase separation, the constant volume formation energies now show ordering/mixing preferences between Si and Ge. 
The purely chemical interactions between Si and Ge at constant volume are therefore slightly attractive. 
Similar mixing formation energies are predicted when fixing the volume to that of pure Si (\Cref{fig:SiGeHullatpureSi}) or to that of pure Ge (\Cref{fig:SiGeHullatpureGe}).

\begin{figure}[htbp]
    \subfloat[]{\includegraphics[width=7cm]{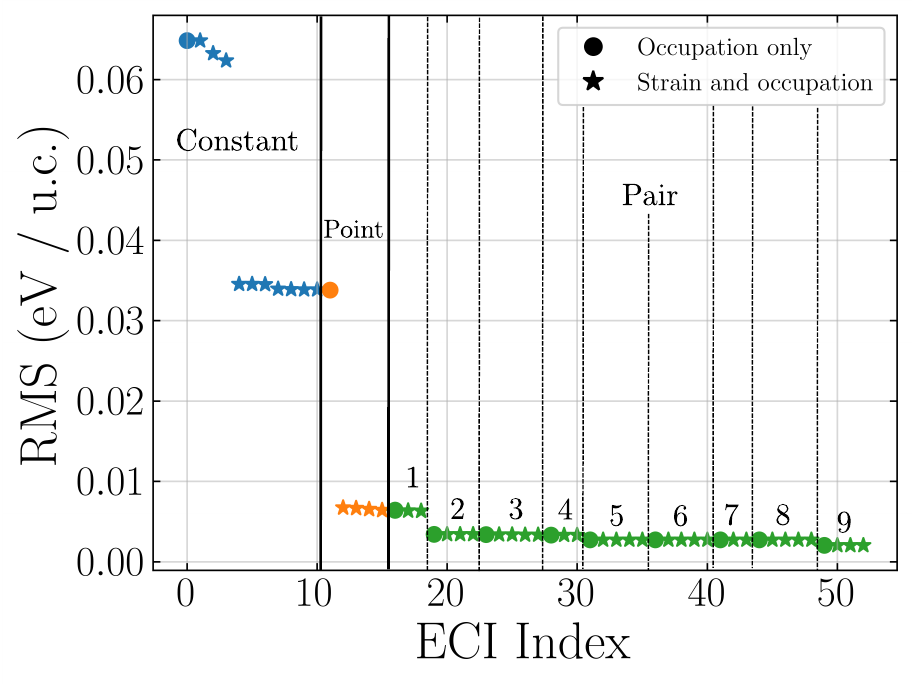}\label{fig:cv_vs_eci}}\\
   \subfloat[]{\includegraphics[width=7cm]{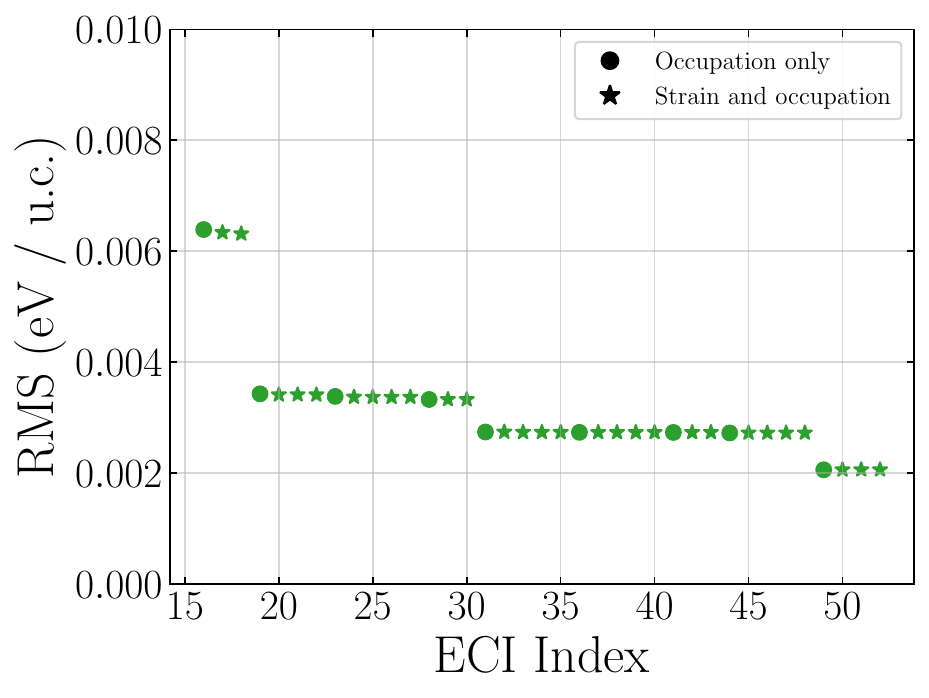}\label{fig:cv_vs_eci_coupled_only_pairs}}\\
    \subfloat[]{\includegraphics[width=7cm]{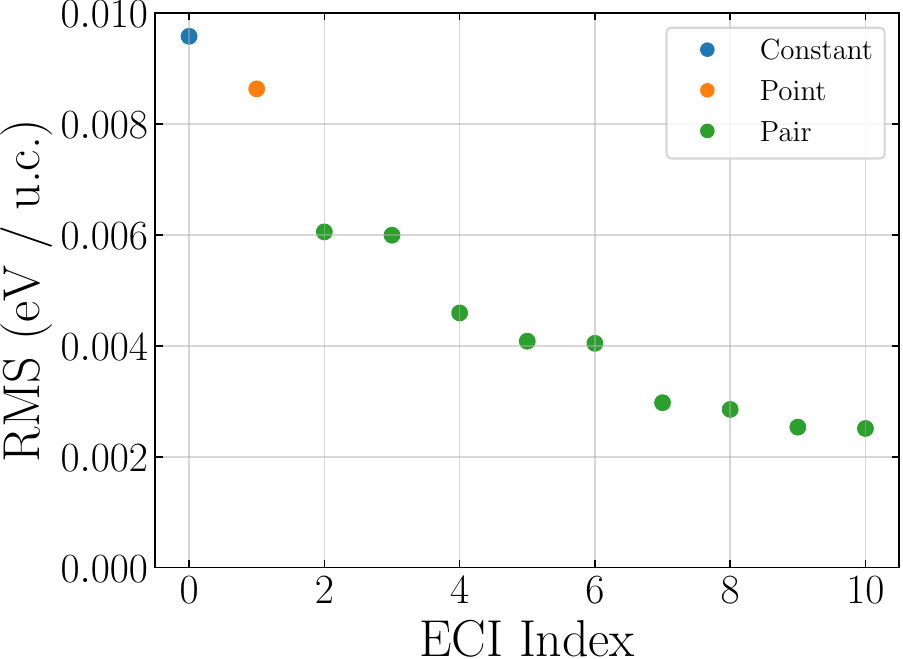}\label{fig:cv_vs_eci_global}}
    {
    \caption{(a) The root mean square error (rmse) upon the incremental addition of cluster basis functions to a truncated configuration plus strain cluster expansion; (b) Same as (a), but within an rmse range of 0 to 10 meV/unit cell; (c) rmse upon the incremental addition of cluster basis functions for a configuration only cluster expansion.}
    \label{fig:fits}
    }
\end{figure}

The results of \Cref{fig:hullsatdifferente1s} show that there is a strong coupling between chemistry and mechanics in the Si-Ge alloy. 
To treat this coupling explicitly, we parameterized a strain plus configuration cluster expansion by training to the energies of a large number of systematically strained Si-Ge orderings. 
For each of the 119 orderings, a grid of symmetrically distinct strain states were enumerated within the $e_1$ irreducible subspace, within the $e_2$-$e_3$ irreducible subspace at each $e_1$ strain and within the $e_4$, $e_5$ and $e_6$ irreducible subspace at each $e_1$ and $e_2$-$e_3$ strains. 
The energies of a total of 14395 chemical plus strain configurations were calculated, allowing only relaxations of internal atomic coordinates. 
A compact configuration plus strain cluster expansion was then fit to these configurations.
The configuration plus strain cluster expansion was able to reproduce the training data with high fidelity, having a root mean squared error of 0.0021 eV per unit cell (or 0.001 eV per atom). Only 53 ECI are needed in a strain plus configuration cluster expansion to achieve this accuracy. 
\Cref{fig:cv_vs_eci} and \Cref{fig:cv_vs_eci_coupled_only_pairs} shows the rmse of the cluster expansion changes when the next basis function in the expansion is added as part of the truncated cluster expansion. An rmse of 0.0035 eV per unit cell can be achieved with only 23 terms in the cluster expansion (these are terms that extend up to the second nearest neighbor).

The resultant strain plus configuration cluster expansion was tested on the fully relaxed structures. 
Each fully relaxed structure has a particular chemical ordering, $\vec{\sigma}$, and a relaxation strain, $\vec{e}$, relative to an initial reference state. 
Inserting these values into the strain plus configuration cluster expansion makes it possible to predict the energy of the fully relaxed state. 
We find that the strain plus configuration cluster expansion is capable of reproducing the fully relaxed energies (to which it was not explicitly trained) with an rmse of 0.0025 eV per unit cell (0.001 eV per atom).

To compare the strain plus configuration cluster expansion to a traditional alloy cluster expansion, we also parameterized a configuration only cluster expansion by training to the fully relaxed energies. 
A simple least squares method was used to fit a configuration only cluster expansion to the fully relaxed formation energies of 119 Si-Ge orderings (\Cref{fig:SiGe_relaxedhull}). 
A constant, point, and nine pair interactions were sufficient to achieve an rmse of 0.0025 eV per unit cell. \Cref{fig:cv_vs_eci_global} shows how the rmse of a configuration only cluster expansion changes when basis functions of successively larger clusters are added to the truncated cluster expansion. 
While a coupled strain and occupational cluster expansion reached an accuracy of 0.0035 eV per unit cell by only including terms that extend up to the second nearest neighbor, the configuration only cluster expansion requires interactions that extend to at least the fifth nearest neighbor to achieve a similar accuracy. 
This is because the configuration only cluster expansion has to learn not only the chemical interactions among Si and Ge, but also the implicit dependence of the relaxation strain energy on the arrangement of Si and Ge.

\subsection{Chemo-mechanical free energies}

\begin{figure*}[htbp]
    \subfloat[]{\includegraphics[width=8cm]{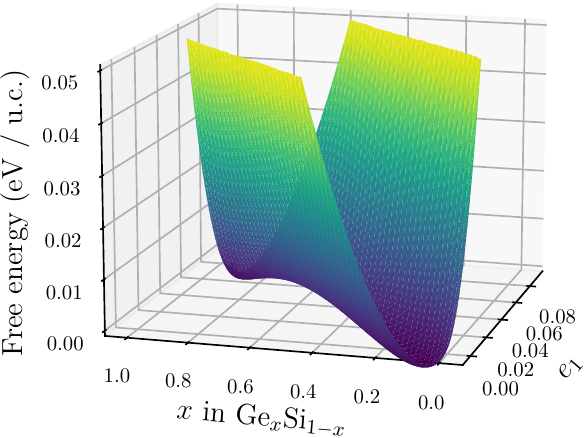}\label{fig:e1xsurface}\hspace{8pt}}
    \subfloat[]{\includegraphics[width=8cm]{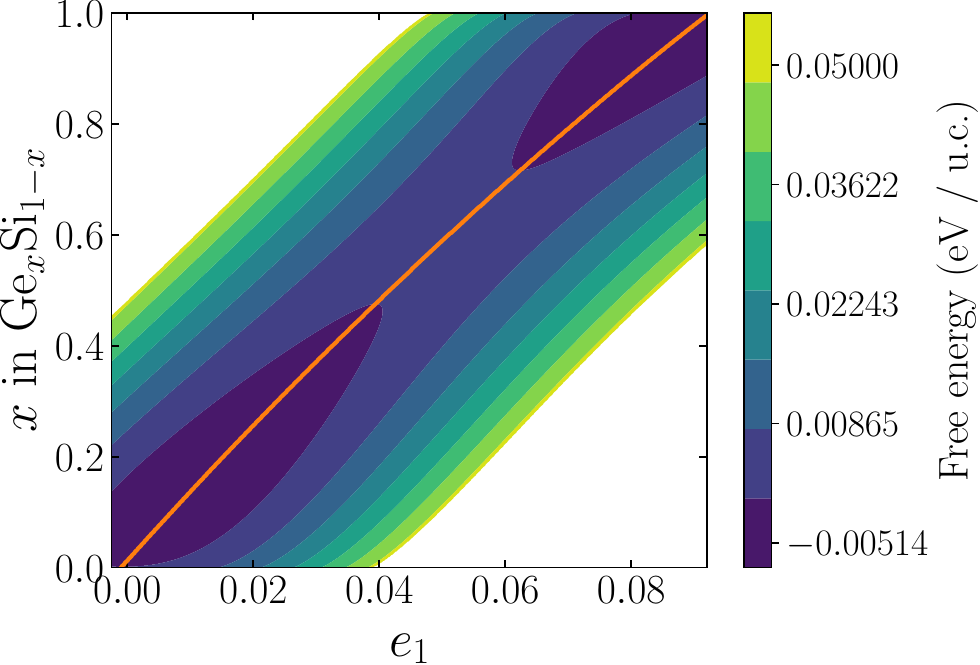}\label{fig:e1xcontour}\hspace{5pt}}\\
    \subfloat[]{\includegraphics[width=8cm]{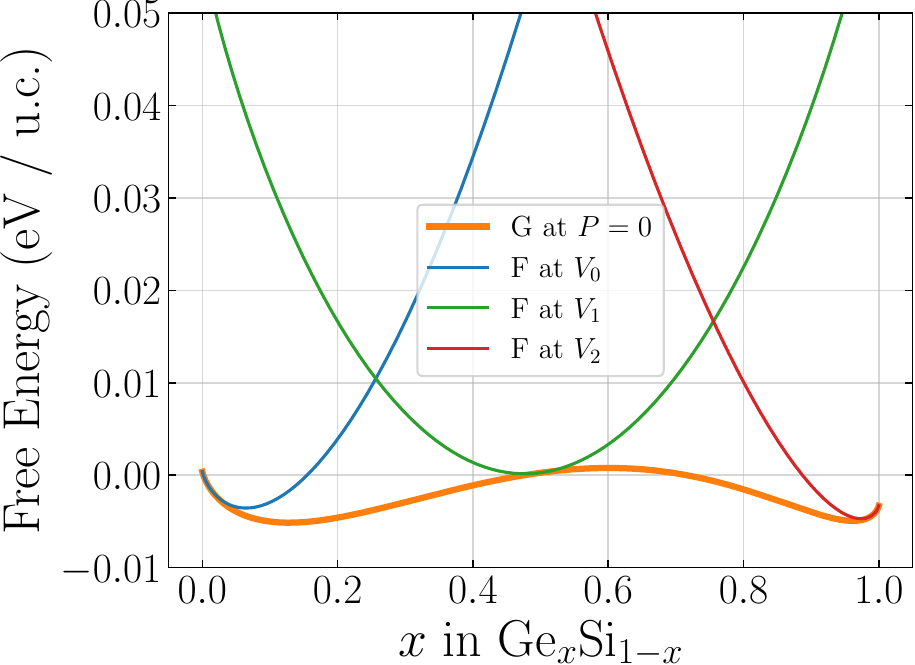}\label{fig:f_x_e1slices}}\hspace{8pt}
    \subfloat[]{\includegraphics[width=8cm]{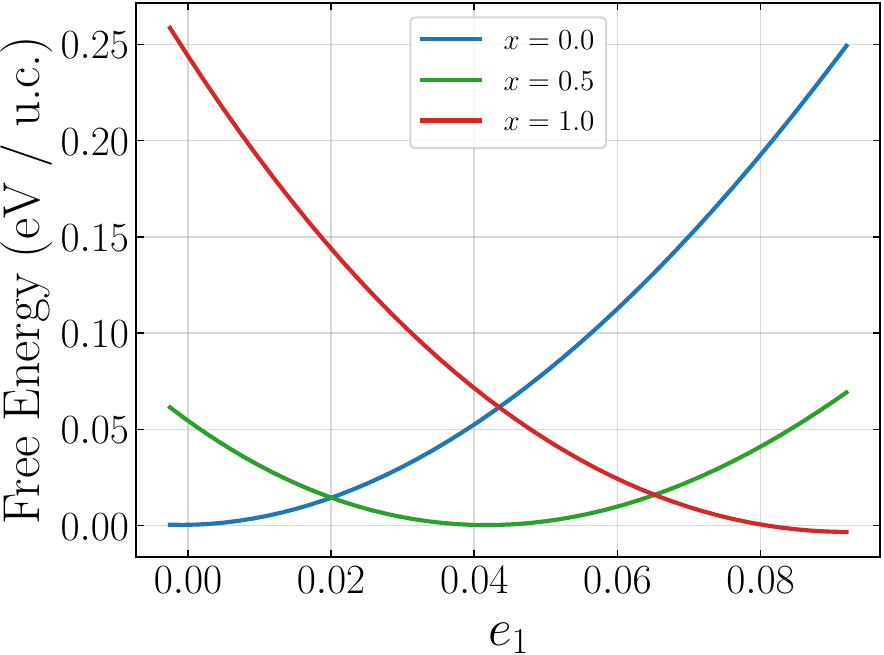}\label{fig:f_e1_xslices}}
    {
    \caption{Helmholtz free energy as a function of concentration $x$ and volumetric strain $e_1$ calculated at 298 K. (a) A three dimensional depiction of the free energy surface and (b) a contour plot. The solid orange line in (b) traces the minimum free energy path in $x$-$e_1$ space, while the minimum free energy at each composition is plotted as a function of concentration in orange in (c). (c) Free energy slices at different values of $e_1$ as a function of composition (d) Free energy slices at different values of composition as a function of $e_1$.}  
    \label{fig:mc}
    }
\end{figure*}

The strain plus configuration cluster expansion was used within grand canonical Monte Carlo simulations to calculate the thermodynamic properties of the Si$_{1-x}$Ge$_x$ alloy as a function of composition, strain, and temperature. 
By holding the strain, $\vec{e}$, in the strain plus configuration cluster expansion constant within the Monte Carlo simulations, it is possible to calculate the Helmholtz free energy as a function of temperature and alloy composition using conventional free energy integration techniques \cite{van2002self,puchala2023casmmonte}.
By repeating this over a grid of strains, a Helmholtz free energy per unit cell, $f(T,x,\vec{e})$, as a function of temperature, $T$, alloy composition, $x$, and strain, $\vec{e}$, can be assembled.

\Cref{fig:e1xsurface} shows the calculated Helmholtz free energy at 298 K as a function of alloy composition $x$ and the $e_1$ strain order parameter, which is a measure of a symmetry preserving volume change. 
The free energy surface has two minima, one at Si rich compositions and low values of $e_1$ and another at Ge rich compositions and high values of $e_1$. 
\Cref{fig:e1xcontour} shows a contour plot of the free energy surface as a function of $x$ and $e_1$. 
The solid orange curve in \Cref{fig:e1xcontour} traces out the minimum in the free energy with respect to $e_1$ for every fixed concentration $x$, i.e. $e_1(x)$ that satisfies $(\partial f/\partial e_1)_{x,T}=0$. 
This is the value of $e_1$ at which the solid has a zero pressure, as can be seen from the thermodynamic equation of state relationship
\begin{equation}
    P=-\left(\frac{\partial F}{\partial V}\right)_{N_{Si},N_{Ge},T}=-\left(\frac{\partial f}{\partial e_1}\right)_{x,T}\left(\frac{\partial e_1}{\partial v}\right)
\end{equation}
where $F$ is the free energy of the solid, $V$ is the volume and $N_{Si}$ and $N_{Ge}$ are the number of Si and Ge atoms, respectively. 
For a solid consisting of $N_u$ unit cells, $F=N_{u}f$ and $V=N_{u}v$, where $v$ is the volume per unit cell. 
The other strain order parameters, $e_2,\dots,e_6$ are implicitly held constant in all the partial derivatives.

Since atmospheric pressure is very close to zero relative to typical elastic stresses in the solid state, it is common in the {\it ab initio} literature to approximate equilibrium at atmospheric pressure as equivalent to equilibrium at zero pressure. 
The Helmholtz free energy along $e_1(x)$ satisfying $(\partial f/\partial e_1)_{x,T}=0$ then also coincides with the Gibbs free energy of the solid at zero pressure since $G=F+PV$, which simplifies to $G=F$ when $P=0$.
The orange curve in \Cref{fig:f_x_e1slices} plots this free energy as a function of composition $x$. 
It shows the characteristic double-well free energy curve that produces a miscibility gap in a temperature versus composition phase diagram. 
However, this free energy is the unconstrained free energy, coinciding with the zero-pressure equilibrium volume. 
The miscibility gap can only emerge for an incoherent two-phase mixture, where the coexisting phases are free to relax to their equilibrium volumes at zero pressure. 
Also shown in \Cref{fig:f_x_e1slices} are three free energy curves as a function of composition, but taken at constant slices of $e_1$. 
These free energies have positive curvatures and therefore exhibit a tendency for chemical mixing (as opposed to phase separating). 
Any attempt to realize a coherent phase separation when the solid is constrained to a fixed value of $e_1$ results in an increase in the free energy. 
In the presence of coherency constraints, the Si-Ge alloy will therefore form a disordered solid solution. 
Each of these curves is tangent to the zero pressure free energy curve at the composition $x$ coinciding with the imposed $e_1$.

\begin{figure}[htbp]
    \centering
    \includegraphics[width=8cm]{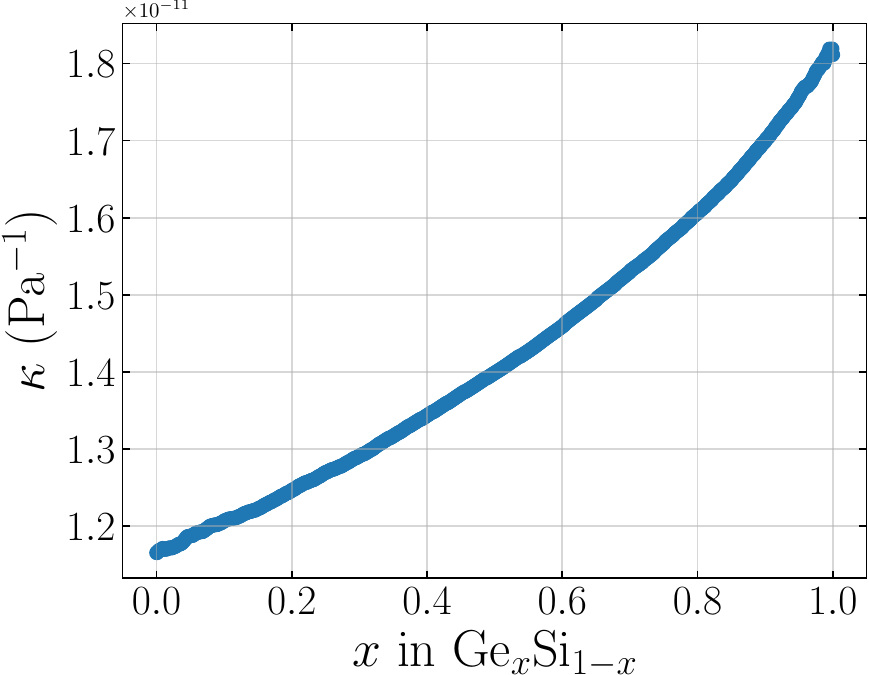}
    \caption{The calculated compressibility of the Ge$_x$Si$_{1-x}$ alloy as a function of concentration at $P=0$.}
    \label{fig:compressibility}
\end{figure}

\Cref{fig:f_e1_xslices} shows free energy slices at three fixed concentrations plotted as a function of $e_1$. 
The curvature of the free energy with respect to $e_1$ at fixed concentration is related to the compressibility of the solid 
\begin{equation}
    \kappa = -\frac{1}{v}\left(\frac{\partial v}{\partial P}\right)=\frac{1}{v}\frac{1}{\left(\frac{\partial^2f}{\partial v^2}\right)}
\end{equation}
where $T$, $x$, $N_u$ (number of unit cells) and the strains, $e_2,\dots,e_6$ are held constant in the above derivatives. 
The second derivative of the free energy, $f$, with respect to volume can be expressed in terms of derivatives with respect to $e_1$ according to
\begin{equation}
    \left(\frac{\partial^2 f}{\partial v^2}\right) = \left[\left(\frac{\partial^2 f}{\partial e_1^2}\right)\left(\frac{\partial e_1}{\partial v}\right)^2 + \left(\frac{\partial f}{\partial e_1}\right)\left(\frac{\partial^2 e_1}{\partial v^2}\right)\right]
\end{equation}
where $T$, $x$, and the strains, $e_2,\dots,e_6$ are again held constant. 
\Cref{fig:compressibility} shows the dependence of the compressibility of Ge$_x$Si$_{1-x}$ as a function of alloy concentration at 298 K. 
The compressibility of the alloy increases almost linearly between the compressibilities of Si and Ge. 
The calculated compressibility of pure Si of $1.16 \times 10^{-11}$ \si{\per\pascal} is in good agreement with the experimentally measured compressibility of $1.018 \times 10^{-11}$ \si{\per\pascal} \cite{bartl2020thermal}. 
The agreement between the calculated compressibility for pure Ge of $1.81 \times 10^{-11}$ \si{\per\pascal} and the experimentally reported value of $1.33 \times 10^{-11}$ \si{\per\pascal} \cite{mcskimin1958elastic} is not as good as for pure Si. 
The difference, however, can be attributed to the choice of the exchange-correlation functional in our DFT calculations. 
Previous \textit{Ab initio} studies using GGA PBE report similar values for the Ge compressibility ($1.78 \times 10^{-11}$\si{\per\pascal}) \cite{wang2003first}. 
The full tensor of elastic moduli of the Ge$_x$Si$_{1-x}$ alloy are related in a similar way to second derivatives of the chemo-mechanical free energy $f(T,x,\vec{e})$ with respect to the six independent Cartesian strains $E_{xx}$, $E_{yy}$, $E_{zz}$, $E_{xy}$, $E_{xz}$ and $E_{yz}$.
The mathematical relationship between the elastic moduli and the second derivatives of the chemo-mechanical free energies depends on the type of strain that is used (i.e. Green-Lagrange or a Hencky strain).

\begin{figure*}[htbp]
    \subfloat[]{\includegraphics[width=8cm]{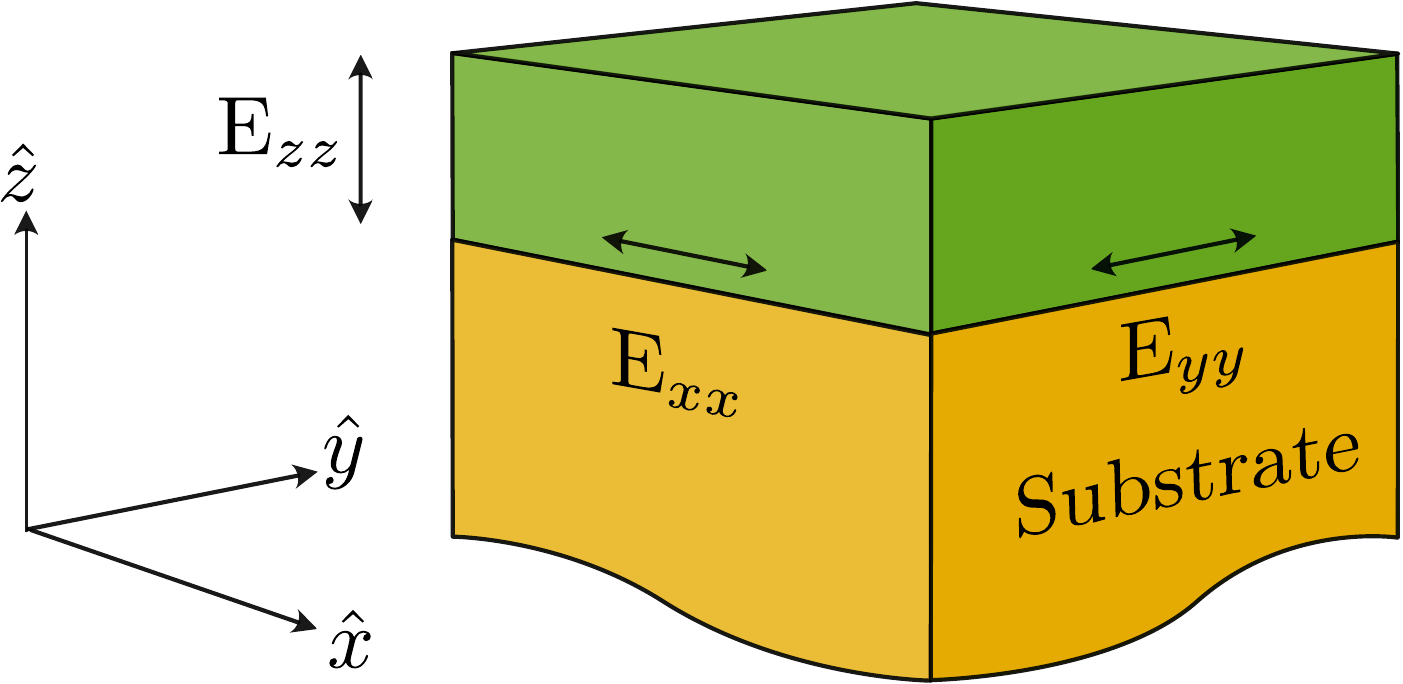}
    \label{fig:Epitaxy}}\hspace{5pt}
    \subfloat[]{\includegraphics[width=8cm]{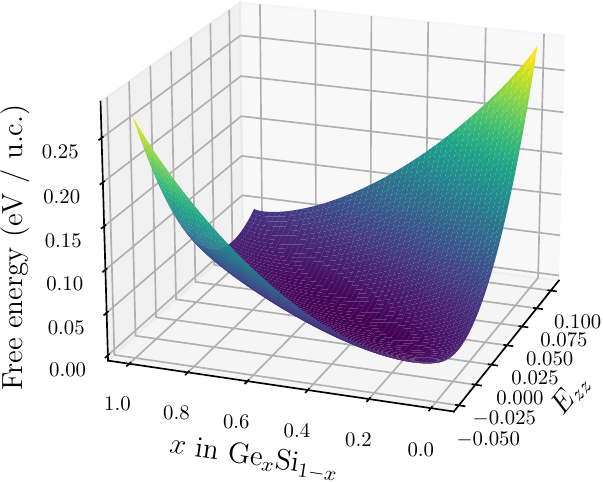}\label{fig:ezz_energy_surface}}\\
    \subfloat[]{\includegraphics[width=7.5cm]{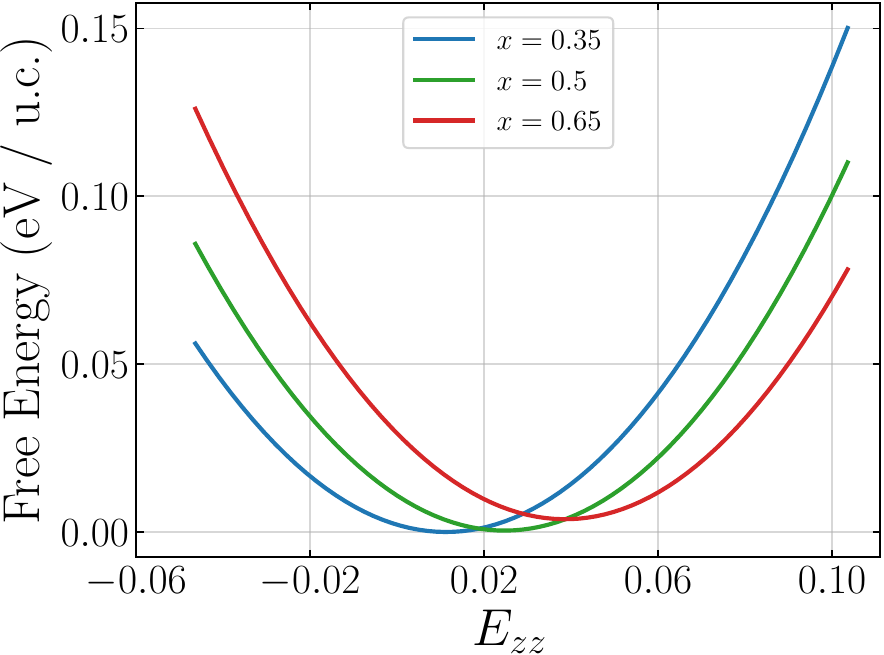}\label{fig:f_ezz_xslices}} \hspace{5pt}
   \subfloat[]{\includegraphics[width=8cm]{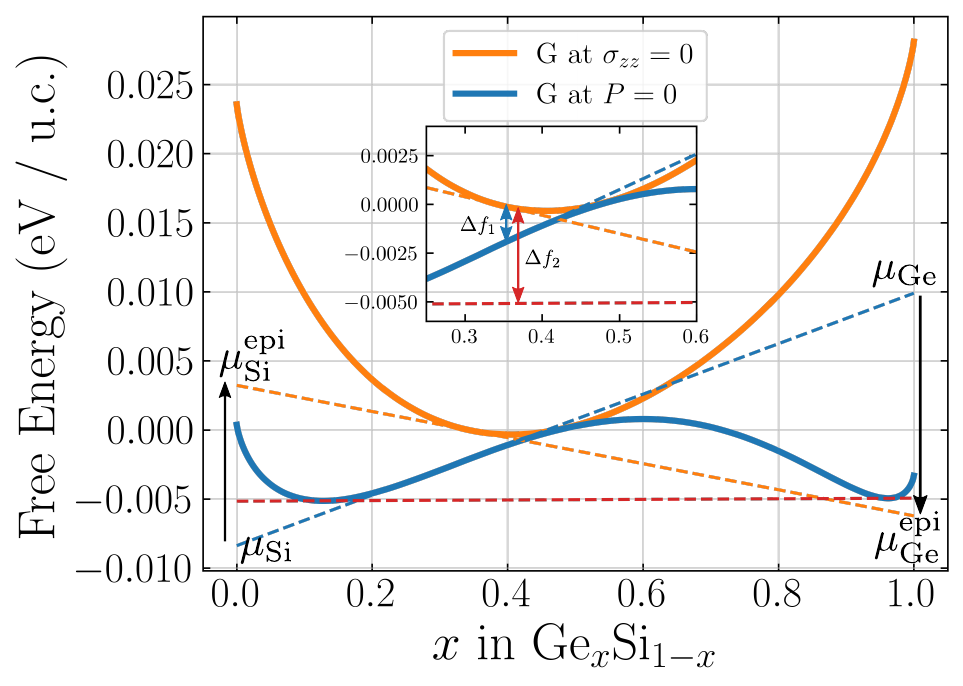}\label{fig:G_ezz_e1}}
    {
    \caption{ (a) Schematic illustration of epitaxial growth; (b) Helmholtz free energy at room temperature calculated as a function of composition and $E_{zz}$ under epitaxial strain conditions (fixed $E_{xx}$ and $E_{yy}$) (b) Free energy slices at $x=0.35, 0.5, 0.65$ as a function of $E_{zz}$. (c) The solid orange line shows the epitaxially constrained free energy coinciding with $\sigma_{zz} = 0$. The solid blue line corresponds to the equilibrium free energy when $P=0$. The dotted orange line shows the tangent to the epitaxially constrained free energy at $x=0.35$. The black arrows indicate how the chemical potentials of Si and Ge change upon going from an unconstrained state to an epitaxially strained state at fixed $E_{xx}$ and $E_{yy}$ and $\sigma_{zz}=0$.}
    \label{fig:mc_epitaxial} 
    }
\end{figure*}

A simple example where chemo-mechanical free energies are important is in the context of coherent epitaxial growth of alloys on a "semi-infinite" substrate (Figure \ref{fig:Epitaxy}). 
As a concrete example, we consider a Ge$_x$Si$_{1-x}$ alloy (green in Figure \ref{fig:Epitaxy}) grown on the (001) surface of a Ge$_{0.5}$Si$_{0.5}$ substrate (yellow in Figure \ref{fig:Epitaxy}). 
If the epitaxially grown alloy has a composition that differs from that of the substrate, it will be subjected to epitaxial strains that alter its free energy. 
The semi-infinite substrate constrains the dimensions of the epitaxial alloy in the $\hat{x}-\hat{y}$ plane, which is parallel to the substrate surface. 
The substrate therefore fixes the  $E_{xx}$, $E_{yy}$ and $E_{xy}$ strains of the epitaxial alloy to the equilibrium values of the Ge$_{0.5}$Si$_{0.5}$ substrate alloy. 
Perpendicular to the substrate, however, the epitaxial alloy is free to relax along the $\hat{z}$ direction until mechanical equilibrium is reached with the externally imposed stress, which is usually determined by the pressure, $P$, of the gaseous environment. 
For the simple geometry of this example, the stress within the epitaxial alloy along the $\hat{z}$ direction is related to the first derivative of the free energy  with respect to the Hencky strain $E_{zz}$ according to
\begin{equation}
    \sigma_{zz}=\frac{1}{v}\frac{\partial f}{\partial E_{zz}}=\frac{1}{v}\left[\frac{\partial f}{\partial e_1}\frac{\partial e_1}{\partial E_{zz}}+\frac{\partial f}{\partial e_3}\frac{\partial e_3}{\partial E_{zz}} \right].
    \label{eq:stress}
\end{equation}
In this expression, $v=V/N_u$ is the volume per unit cell in the current state.
All derivatives are taken while holding $T$, $N_{Si}$, $N_{Ge}$ and the strains $E_{xx}$, $E_{yy}$, $E_{yz}$, $E_{xz}$ and $E_{xy}$ constant.

\Cref{fig:ezz_energy_surface} shows the calculated chemo-mechanical free energy as a function of alloy concentration and $E_{zz}$ at the constant strains $E_{xx}$ and $E_{yy}$ imposed by the substrate ($E_{yz}=E_{xz}=E_{xy}=0$ due to the symmetry of the problem). 
\Cref{fig:f_ezz_xslices} shows slices of the free energy at three different alloy compositions as a function of $E_{zz}$. 
When the film is in equilibrium with a low pressure environment, such as vacuum or atmospheric pressure, $P\approx 0$, and the equilibrium value of $E_{zz}$ can be determined by setting Eq. \ref{eq:stress} to zero. 
This is equivalent to minimizing the chemo-mechanical free energy of \Cref{fig:ezz_energy_surface} with respect to $E_{zz}$ at each alloy composition. 
The resulting minimum free energy is shown as the orange curve in Figure \ref{fig:G_ezz_e1}.
This free energy is compared to the free energy of the alloy at $P=0$, without epitaxial constraints (blue curve). 
The two curves touch each other at $x=0.5$ since the epitaxial alloy then has the same composition as the substrate and there are no epitaxial strains. 
When the composition of the epitaxial alloy differs from that of the substrate, the epitaxial misfit strains in the $\hat{x}$ and $\hat{y}$ directions result in a strain energy cost, which raises the free energy of the epitaxial alloy relative to an unconstrained alloy at the same composition. 
Hence the free energy of the epitaxial alloy (orange curve) is always above that of the unconstrained alloy at $P=0$ (blue curve).

The chemical potentials of Si and Ge can be determined graphically from the intercept of the tangent to the free energy curves with the energy axis at $x=0$ and $x=1$, respectively, as shown in Figure \ref{fig:G_ezz_e1}. 
The red dashed line in Figure \ref{fig:G_ezz_e1} corresponds to the common tangent to the free energy curve at zero pressure, which predicts an incoherent miscibility gap at zero pressure. 
The orange dashed line is tangent to the free energy of the epitaxially strained alloy at $x=0.35$. 
Its intercepts with the free energy axis at $x=0$ and $x=1$ correspond to the Si and Ge chemical potentials in the epitaxially strained alloy. 
The blue dashed line is tangent to the free energy of the unconstrained alloy at $x=0.35$. 
The Si and Ge chemical potentials of the epitaxially constrained alloy at $x=0.35$ differ from their values in the unconstrained alloy at the same composition $x=0.35$.
Since the epitaxial alloy at $x=0.35$ must undergo a positive strain in the $\hat{x}$ and $\hat{y}$ directions ($E_{xx}>0$ and $E_{yy}>0$) as it is coherently attached to a substrate having a composition $x=0.5$ (which has a larger volume), the chemical potential of the smaller Si atoms increases, while the chemical potential of the larger Ge atoms decreases relative to their values in the unconstrained alloy at the same composition.  
The epitaxial strain energy, therefore, sets up diffusional driving forces that favor the diffusion of Ge (Si) to (away from) the epitaxially strained alloy. 

The difference in the free energy between the epitaxial alloy and the unconstrained alloy at $x=0.35$, denoted $\Delta f_1$ in Figure \ref{fig:G_ezz_e1}, can be interpreted as the strain free energy per unit cell that is needed to elastically deform the unconstrained alloy to achieve epitaxy with the substrate.  
The total strain free energy of the epitaxial alloy scales with the size of the film and therefore will be linearly proportional to the film thickness. 
Above a critical thickness, the strain energy will exceed the cost of introducing misfit dislocations along the substrate/film interface that can relieve the epitaxial constraints and allow the alloy to lower its free energy. 
If the constituents of the epitaxially strained alloy phase separate to form an incoherent two-phase mixture, the free energy will lower by $\Delta f_2$ as illustrated in Figure \ref{fig:G_ezz_e1}.

\section{Discussion}

We have introduced a strain plus configuration cluster expansion formalism that enables a rigorous treatment of chemical and strain degrees of freedom in multi-component crystals. 
The approach makes it possible to calculate free energies that explicitly depend on temperature, composition, and strain. 
The first derivatives of such free energy surfaces determine chemical potentials and stresses, while their curvatures with respect to strain variables are related to elastic moduli.\cite{thomas2014elastic} 
A free energy description that explicitly couples composition with strain, therefore, enables a straightforward calculation of the dependence of the elastic moduli on concentration and short-range order.

The strain variables that appear in the chemo-mechanical free energy are measured relative to a fixed reference state, such as the fully relaxed diamond unit cell of pure Si for the Ge$_x$Si$_{1-x}$ alloy example considered in this work. 
Since the equilibrium volume and shape of the unit cell can change substantially upon varying the alloy composition, it is important to use measures of finite strain that are invariant to rigid rotations. 
The Green-Lagrange strain can serve this purpose, but for first-principles free energy surfaces expressed in terms of symmetry adapted strain order parameters, it is often more convenient to work with the Hencky strain.\cite{goiri2016phase,thomas2017exploration,behara2022ferroelectric} 
This is because the symmetry invariant strain order parameter, $e_1$, is then directly proportional to the volume change, and any variations in the symmetry breaking strain order parameters, $e_2,\dots,e_6$, at constant $e_1$, occur at constant volume.\cite{thomas2017exploration} 
Due to the reliance on the same reference state for strain at all compositions, the strain variables can be separated into a chemical strain, $e_{i}^{chem}(x)$, which depends on the alloy composition, and an elastic strain, $e_{i}^{elast}$, according to
\begin{equation}
    e_{i}=e_{i}^{chem}(x)+e_{i}^{elast}
\end{equation}
The chemical strain corresponds to the strain that minimizes the free energy at each composition. 
The elastic strain then measures deformations of the crystal relative to its equilibrium dimensions at that composition given by $e_i^{chem}(x)$.

Free energy surfaces that explicitly depend on composition and strain are the natural ingredients to phase-field simulations\cite{chen2002phase} of microstructure evolution that couples chemical redistribution through diffusion with elastic driving forces due to coherency strains.\cite{rudraraju2016mechanochemical} 
A phase field model that describes coherent microstructure evolution departs from a free energy of the form
\begin{equation}
    \int_V \left[f(T,x,\vec{e})+\gamma(T,x,\vec{e})\right]\frac{d\vec{r}}{v}
\end{equation}
where the first term of the integrand is the homogeneous free energy and the second term collects gradient energy contributions that correct the homogeneous free energy in the presence of gradients in composition and strain. 
The gradient energy contributions take the form
\begin{equation} \label{eq1}
\begin{split}
\gamma(T,x,\vec{e}) & = \sum_{\alpha,\beta}K_{\alpha,\beta}(x,\vec{e})\frac{\partial x}{\partial r_{\alpha}}\frac{\partial x}{\partial r_{\beta}} \\
 & +\sum_{i,j}\sum_{\alpha,\beta}\Gamma^{i,j}_{\alpha,\beta}(x,\vec{e})\frac{\partial e_i}{\partial r_{\alpha}}\frac{\partial e_j}{\partial r_{\beta}} \\
 & +\sum_i\sum_{\alpha,\beta}\Lambda_{\alpha,\beta}^i(x,\vec{e})\frac{\partial x}{\partial r_{\alpha}}\frac{\partial e_i}{\partial r_{\beta}}
\end{split}
\end{equation}
where $r_{\alpha}$ and $r_{\beta}$ represent components of a Cartesian vector and where $K_{\alpha,\beta}$, $\Gamma^{i,j}_{\alpha,\beta}$ and $\Lambda_{\alpha,\beta}^i$ are gradient energy coefficients. 
The symmetry of the parent crystal imposes constraints between the different gradient energy coefficients.\cite{kitchaev2018phenomenology}

Gradient energy contributions are only strictly necessary if the homogeneous free energy exhibits negative curvatures along particular directions in the variable space spanned by $x$ and $e_1,\dots,e_6$.
The gradient energy term then regularizes the homogeneous free energy in phase-field simulations of spinodal decomposition or ordering reactions.\cite{cahn1958free,cahn1961spinodal,allen1979microscopic}
The calculated chemo-mechanical free energy surfaces of the Ge$_x$Si$_{1-x}$ alloy have positive curvatures with respect to both composition (Figure \ref{fig:f_e1_xslices}) and strain (Figure \ref{fig:f_x_e1slices}). 
It is only along directions that mix composition and the $e_1$ strain that negative curvatures emerge in the chemo-mechanical free energy of Ge$_x$Si$_{1-x}$, as is evident in Figure \ref{fig:e1xcontour}.
A description of spinodal decomposition in the Ge$_x$Si$_{1-x}$ alloy would then require symmetry allowed gradient energy coefficients, $\Lambda_{\alpha,\beta}^i$, that couple gradients in composition with gradients in strain, i.e. $(\partial x/\partial r_{\alpha})(\partial e_i/\partial r_{\beta})$. 
This is an overlooked subtlety in treatments of spinodal decomposition in the presence of coherency strain, where a negative curvature with respect to the composition axis only is generally assumed and gradient energy terms that couple gradients in composition with gradients in strain are neglected.\cite{cahn1958free,cahn1961spinodal,cahn1962coherent}
Some multi-component solids can also exhibit negative curvatures with respect to symmetry breaking strains, allowing for the occurrence of chemo-mechanical spinodal decomposition.\cite{thomas2013finite,thomas2014elastic,rudraraju2016mechanochemical} 
In this case, strain gradient energy contributions must be included in the free energy description.\cite{rudraraju2014three}
The chemo-mechanical free energy can also be extended to include a dependence on order parameters that track chemical orderings over the parent crystal structure.\cite{natarajan2017symmetry,van2018first}

Cluster expansions that only depend on chemical configurational degrees of freedom are unable to accurately treat the long-range effects of coherency strains when the expansion is truncated, instead being most suited for the calculation of the thermodynamic properties of homogeneous phases. 
The mixed basis cluster expansion approach of Laks et al.\cite{laks1992efficient} was designed to rectify this deficiency. 
While a powerful method to model alloy thermodynamics \cite{ozolicnvs1998cu,wolverton1998first,wolverton2000short,liu2008thermodynamic,liu2009thermodynamic,wang2023generalization} and coherent multi-phase equilibrium at the meso scale \cite{wolverton2000first}, the method does not enable the calculation of free energies that explicitly depend on strain. 
Furthermore, it is considerably more complex than a real-space cluster expansion that can be truncated beyond a maximal cluster size, as it requires the calculation of the constituent strain energy and frequent evaluations of Fourier transforms within Monte Carlo simulations.  
Finally, while such a treatment can be expected to be rigorous for systems that exhibit simple miscibility gaps, it becomes more questionable when complex ordered phases that cannot be described as simple layered orderings are stable. 
The strain plus configuration cluster expansion introduced in this work addresses many of the shortcomings of both the real-space and mixed-basis cluster expansions for solids that can undergo large variations in volume and group/subgroup symmetry breaking strains with composition. 
As a surrogate model for first-principles statistical mechanics simulations, the strain plus configuration cluster expansion should be a convenient compromise between the more restrictive configuration only cluster expansion \cite{sanchez1984generalized,de1994cluster} and the more versatile machine-learned interatomic potentials,\cite{drautz2019atomic,musil2021physics,deringer2021gaussian} which treat all degrees of freedom, but require a substantially larger effort to train with first principles electronic structure calculations.

\section{Conclusion}
We have described a statistical mechanics approach to calculate chemo-mechanical free energies that depend on temperature, composition, and strain. 
This is enabled with an extension of the alloy cluster expansion to also include a dependence on homogeneous strain variables. 
The symmetry of the parent crystal structure imposes constraints on the form of the expansion basis functions.
To illustrate the approach, we calculated the chemo-mechanical free energy of the Si-Ge alloy. 
In the absence of coherency constraints at zero pressure, the alloy has phase separating tendencies. 
Coherency constraints, however, modify phase stability and lead to mixing tendencies among Si and Ge on the diamond parent crystal structure. 
Chemo-mechanical free energy surfaces that depend not only on temperature and composition but also explicitly on finite measures of strain are essential ingredients to rigorous phase field models of coherent microstructure evolution.

\section{Acknowledgement}

SSB was supported by the Defense Advanced Research Projects Agency under Contract No. HR001122C0097.
JCT, BP and AVDV acknowledge support from the U.S. Department of Energy, Office of Basic Energy Sciences, Division of Materials Sciences and Engineering under Award $\#$DE-SC0008637 as part of the Center for Predictive Integrated Structural Materials Science (PRISMS Center).
Use was made of computational facilities purchased with funds from the National Science Foundation (CNS-1725797) and administered by the Center for Scientific Computing (CSC). The CSC is supported by the California NanoSystems Institute and the Materials Research Science and Engineering Center (MRSEC; NSF DMR 2308708) at UC Santa Barbara. We are also grateful for the resources of the National Energy Research Scientific Computing Center (NERSC), a U.S. Department of Energy Office of Science User Facility located at Lawrence Berkeley National Laboratory, operated under Contract No. DE-AC02-05CH11231 using NERSC award BES-ERCAP0023147.

\bibliography{./references.bib}
\end{document}